# Quantum Phase Transition of the Twisted Spin Tube

Yuta TACHIBANA[1*], Yuki UENO[1], Tomosuke ZENDA[1], Kiyomi OKAMOTO[1] and Toru SAKAI[1,2]

[1]*Graduate School of Material Science, University of Hyogo, Hyogo 678-1297, Japan.*
[2]*National Institutes for Quantum and Radiological Science and Technology (QST), SPring-8, Hyogo 679-5148, Japan*

*E-mail: redcliff0220@ezweb.ne.jp*

The $S=1/2$ twisted three-leg spin tube with the lattice distortion from the regular triangle to the isosceles one is investigated using the numerical diagonalization of finite-size clusters and the phenomenological renormalization group analysis. It is found that the quantum phase transition occurs from the spin gap phase to another one with respect to this lattice distortion.

## 1. Introduction

The $S=1/2$ three-leg spin tube [1] system is one of interesting low-dimensional quantum spin systems. It exhibits some exotic quantum phenomena because of strong quantum fluctuation and spin frustration. For example, the $S=1/2$ three-leg spin tube has the spin gap which is the energy gap between the singlet ground state and the triplet excited one [2], while the $S=1/2$ three-leg spin ladder is gapless. The previous theoretical work [3] using the numerical exact diagonalization and the density matrix renormalization group (DMRG) calculation on the $S=1/2$ three-leg spin tube with the lattice distortion from the regular triangle to the isosceles one at each triangle unit indicated that the quantum phase transition occurs from the spin gap phase to the gapless Tomonaga-Luttinger liquid one with respect to this lattice distortion.

One of candidate materials of the $S=1/2$ three-leg spin tube is the compound $[(CuCl_2tachH)_3Cl]Cl_2$ [4]. However, it has a twisted structure and the magnetization measurement indicated that it is gapless. The previous theoretical study [5] on the $S=1/2$ twisted three-leg spin tube using the DMRG calculation revealed that the system can also have the spin gap when the ratio of the inter-triangle and intra-triangle interactions is smaller than the critical value (~1.2). Thus if the lattice distortion from the regular to isosceles triangles is introduced to the spin gap phase, even the $S=1/2$ twisted three-leg spin tube would be expected to exhibit the quantum phase transition from the gapped to gapless phases. In this paper the effect of the lattice distortion on the $S=1/2$ twisted three-leg spin tube is investigated using the numerical exact diagonalization of finite-size clusters and the phenomenological renormalization group analysis in order to detect the quantum phase transition.

## 2. Model

The $S=1/2$ twisted three-leg spin tube with the lattice distortion from the regular to isosceles triangles is described by the Hamiltonian

$$H = J' \sum_{i=1}^{3} \sum_{j=1}^{L} (S_{i,j} \cdot S_{i,j+1} + S_{i,j} \cdot S_{i+1,j+1}) + J \sum_{i=1}^{2} \sum_{j=1}^{L} (S_{i,j} \cdot S_{i+1,j}) + J_1 \sum_{j=1}^{L} (S_{3,j} \cdot S_{1,j})$$

where $S_{i,j}$ is the spin-1/2 operator and $J, J', J_1$ are the exchange coupling constants. The schematic picture of this model is shown in Fig. 1. The ratio $J_1/J$ is defined as $\alpha = J_1/J$.

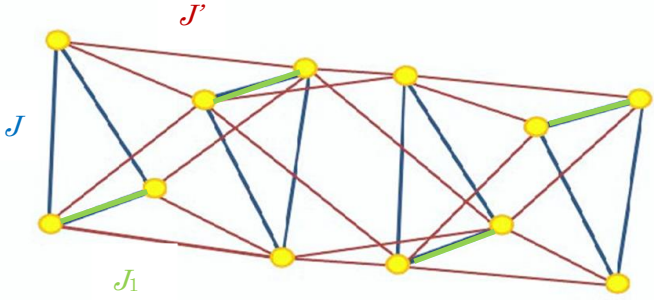

**Fig. 1.** $S=1/2$ twisted three-leg spin tube with the lattice distortion from the regular triangle to the isosceles one.

Each triangle unit is the regular triangle for $\alpha=1$, while the isosceles one for $\alpha \neq 1$. In this paper a possible quantum phase transition is investigated, as the parameter $\alpha$ is varied, using the numerical exact diagonalization of finite-size clusters for $L=4$, 6 and 8. $L$ is the number of the unit cells. Throughout the paper we fix $J=1$. Based on the Lanczos algorithm, we calculate the singlet ground state energy and the triplet excited state energy and estimate the spin gap $\Delta_L$ for the system size $L$. Although the S=1/2 straight three-leg spin tube is trivially gapless for $\alpha=0$ and $\infty$ [3], it is not trivial for the present model, because it is still frustrated even in such limits.

## 3. Quantum phase transition

In order to detect the quantum phase transition the phenomenological renormalization group analysis is one of good methods. At least this method successfully determined the phase boundary between the gapless and gapped phases of the $S=1/2$ straight three-leg tube for $J'<J$ [3]. Now we consider the spin gap $\Delta_L(\alpha)$ as the function of the parameter $\alpha$ with fixed $J'$. According to the phenomenological renormalization group the size-dependent critical point $\alpha_c$ is estimated by the fixed point equation

$(L+2)\Delta_{L+2}(\alpha) = L\Delta_L(\alpha)$

for the two system sizes $L$ and $L+2$. The $\alpha$ dependence of the scaled gap $L\Delta_L(\alpha)$ with fixed

$J'=0.2$ and $0.4$ is shown in Figs. 2 and 3, respectively. In both cases it is found that the two quantum phase transitions where the spin gap vanishes occur at $\alpha_{c1}<1$ and $\alpha_{c2}>1$. We estimate the size-dependent fixed points $\alpha_{c1}$ and $\alpha_{c2}$ as the cross points for $L=4, 6$ and for $L=6, 8$. Using the estimated critical points, the phase diagram in the $J'$-$\alpha$ plane is obtained as shown in Fig. 4. For smaller $J'$ the size dependence of $\alpha_{c1}$ and $\alpha_{c2}$ is so small that the present phase boundaries are expected to be close to the thermodynamic limit. However, for $J'\sim 1$ the size dependence is too large to determine the phase boundaries in the thermodynamic limit, because the system is close to the gapless phase even for $\alpha=1$ ($J'_c\sim 1.2$).

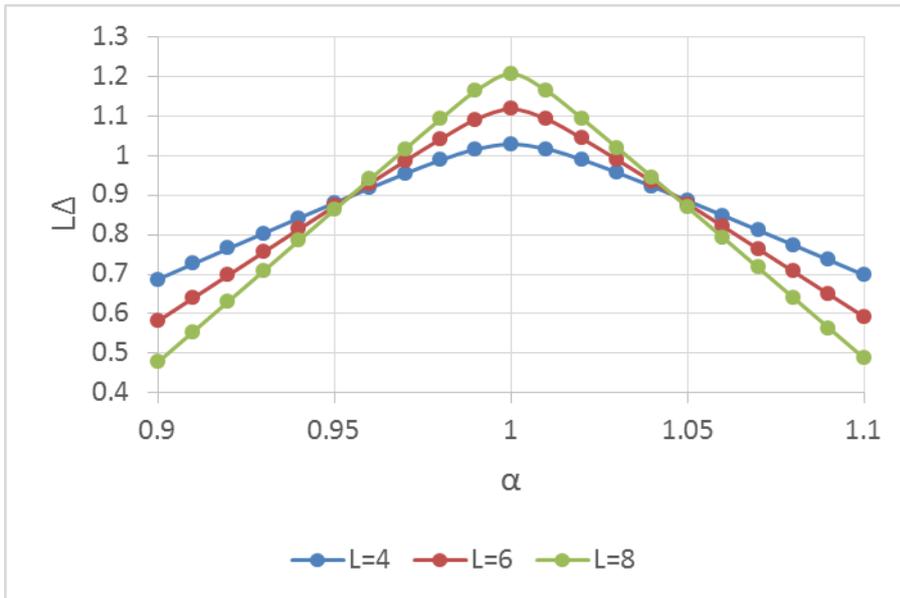

**Fig. 2.** α dependence of the scaled gap $L\Delta$ for $J'=0.2$. It is found that the quantum phase transitions where the spin gap vanishes occur at $\alpha_{c1}\sim 0.95$ and $\alpha_{c2}\sim 1.05$ for $J'=0.2$.

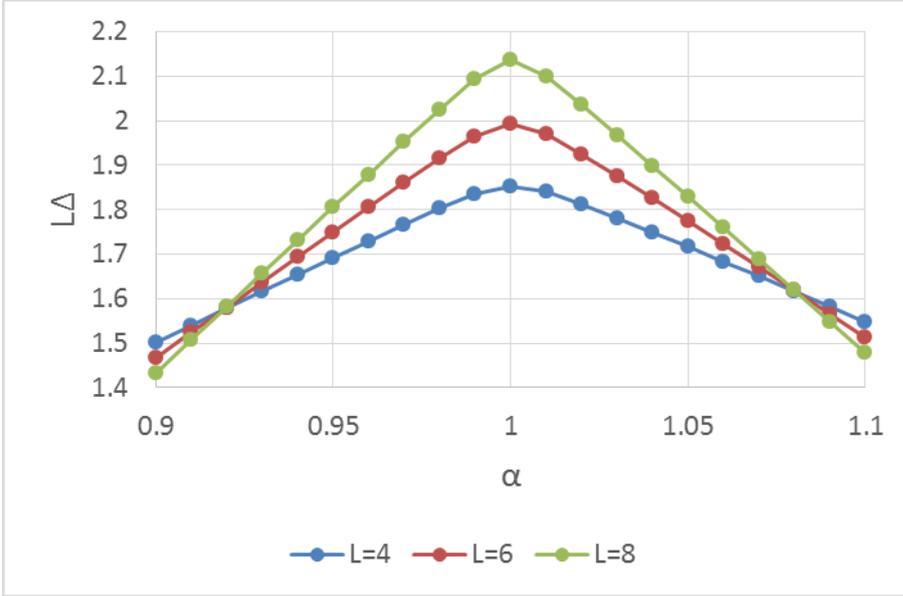

**Fig. 3.** α dependence of the scaled gap $L\Delta$ for $J'$=0.4. It is found that the quantum phase transitions where the spin gap vanishes occur at $\alpha_{c1}$~0.92 and $\alpha_{c2}$~1.08 for $J'$=0.4.

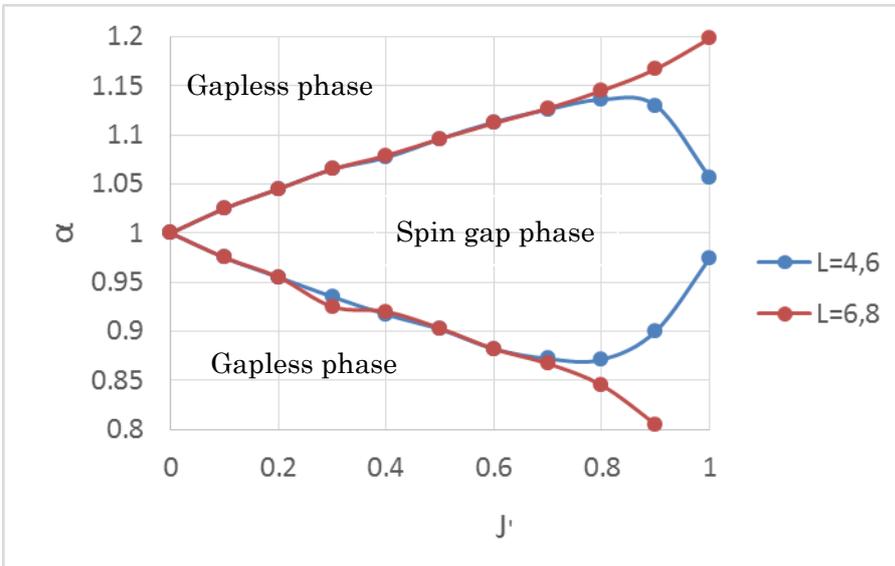

**Fig. 4.** Phase diagram in the $J'$-α plane. Blue (red) curves are the size-dependent phase boundaries determined for $L$=4 and 6 (6 and 8). For $J'$<0.7 the size dependence is so small that the boundaries are supposed to be almost the ones in the thermodynamic limit. For $J'$>0.7 the size dependence is too large to determine the phase boundary in the infinite $L$ limit.

The present phenomenological renormalization group analysis successfully indicates that the quantum phase transition occurs with respect to α and the spin gap vanishes at the critical lines. However, within the present study it is difficult to determine what kind of phases appear. In order to determine it some other methods would be necessary.

4. Summary

The $S$=1/2 twisted three-leg spin tube with the lattice distortion from the regular triangle to the isosceles one is investigated using the numerical exact diagonalization and the phenomenological renormalization group analysis. As a result it is found that the quantum phase transition occurs from the spin gap phase to gapless phases with respect to the distortion. The phase diagram is obtained.

**Acknowledgment**

This work was partly supported by JSPS KAKENHI, Grant Numbers 16K05419, 16H01080 (J-Physics) and 18H04330 (J-Physics). A part of the computations was performed using facilities of the Supercomputer Center, Institute for Solid State Physics, University of Tokyo, and the Computer Room, Yukawa Institute for Theoretical Physics, Kyoto University.